\documentclass{Interspeech2024}

\usepackage{bm}
\usepackage{siunitx}
\usepackage{stfloats}
\usepackage{tipa}
\usepackage{hyperref}

\interspeechcameraready

\title{Towards a Quantitative Analysis of Coarticulation with a Phoneme-to-Articulatory Model}

\name[affiliation={1}]{Chaofei}{Fan}
\name[affiliation={1}]{Jaimie M.}{Henderson}
\name[affiliation={1}]{Chris}{Manning}
\name[affiliation={1,2}]{Francis R.}{Willett}

\address{
  $^1$Stanford University, USA\\
  $^2$Howard Hughes Medical Institute at Stanford University, USA 
}
\email{stfan@stanford.edu}


\begin{document}

\maketitle

\begin{abstract}
    

Prior coarticulation studies focus mainly on limited phonemic sequences and specific articulators, providing only approximate descriptions of the temporal extent and magnitude of coarticulation.
This paper is an initial attempt to comprehensively investigate coarticulation. We leverage existing Electromagnetic Articulography (EMA) datasets to develop and train a phoneme-to-articulatory (P2A) model that can generate realistic EMA for novel phoneme sequences and replicate known coarticulation patterns. We use model-generated EMA on 9K minimal word pairs to analyze coarticulation magnitude and extent up to eight phonemes from the coarticulation trigger, and compare coarticulation resistance across different consonants. Our findings align with earlier studies and suggest a longer-range coarticulation effect than previously found. This model-based approach can potentially compare coarticulation between adults and children and across languages, offering new insights into speech production.

\end{abstract}

\section{Introduction}

Coarticulation is the phenomenon in which one phoneme influences the acoustic and articulatory characteristics of its neighbors.
Understanding coarticulation is crucial for understanding the mechanisms of speech production, the planning and coordination of speech articulators, and how these processes relate to speech disorders and speech acquisition in children \cite{recasens2018coarticulation}.

The analysis of coarticulation has primarily relied on acoustic and articulatory recordings.
However, the challenges in data collection have often limited these analyses to simple phonemic sequences (e.g., consonant-vowel, vowel-consonant-vowel), a narrow selection of phonemes, and specific articulatory movements or spectral formant comparisons \cite{ohman1966coarticulation, fowler2000coarticulation, liu2022coarticulation}.
Questions remain about the full temporal scope of coarticulation across all phonemes and how its degree and impact might differ among speakers and across languages \cite{grosvald2009interspeaker,beddor2002language}.

One potential solution is to use a speech production model to generate accurate synthetic data for any desired experimental condition. 
Formal speech production models, such as DIVA \cite{guenther2016neural} and TADA \cite{nam2004tada}, are built on years of observation but are primarily used to produce articulatory kinematics for short phonemic sequences.
Recent data-driven phoneme-to-articulatory (P2A) and acoustic-to-articulatory (A2A) models \cite{wu23_icassp,biasutto2018phoneme} can accurately produce EMA trajectories for long sentences, but they do not provide the timing of phonemes.

Our first contribution is to develop a data-driven P2A model that can make realistic predictions for what an ``average'' EMA time series would be for any given phoneme sequence, while also returning the time at which each phoneme occurs. 
With this model, we can sample EMA trajectories for any word and then interrogate the model outputs for any coarticulation effect of interest.
Our second contribution is a metric for quantifying the temporal extent and magnitude of coarticulation across all phonemes and articulators.
We use minimal word pairs (two words that differ in only one phonological element) to measure how coarticulation spreads from the differing phoneme (trigger) to its neighboring phonemes (targets).
Compared to previous studies quantifying coarticulation \cite{iskarous2013coarticulation,lindblom2012dissecting,recasens2009articulatory,jackson2009statistical,gelfer1989determining}, ours samples a large number of phonemic sequences and leverages the statistical power of 6+ hours of EMA training data from 12 speakers incorporated by the P2A model.

We train the P2A model on the USC-TIMIT \cite{narayanan2014real} and Haskins Production Rate Comparison \cite{tiede2017quantifying} EMA datasets.
First, we show that the model can generate realistic EMA trajectories, comparable to state-of-the-art (SOTA) A2A models \cite{wu23_icassp,parrot20_interspeech}.
Second, using the model, we assess the effect of coarticulation by generating EMA trajectories on 9,291 minimal word pairs.
Coarticulation effects are most significant for the immediate left and right neighbors of a triggering phoneme ($\pm$1 phonemic distance), and diminish almost threefold at $\pm$2 distance.
Coarticulation effects gradually diminish thereafter but are still significant at $\pm$7 distance.
Coarticulation resistance varies across different places of articulation in consonants. 
Dentals, alveolars, and postalveolars are more coarticulation resistant than bilabials, labiodentals, and velars.
These findings generally align with earlier studies \cite{recasens2009articulatory}, but also suggest a longer-range effect than has been typically found \cite{grosvald2009interspeaker,recasens1989long,magen1997extent}, which might relate to the brain's speech planning mechanism \cite{whalen1990coarticulation,guenther2016neural}.



\section{Phoneme-to-articulatory Model}

\begin{figure}[ht]
  \centering
  \includegraphics[width=\linewidth]{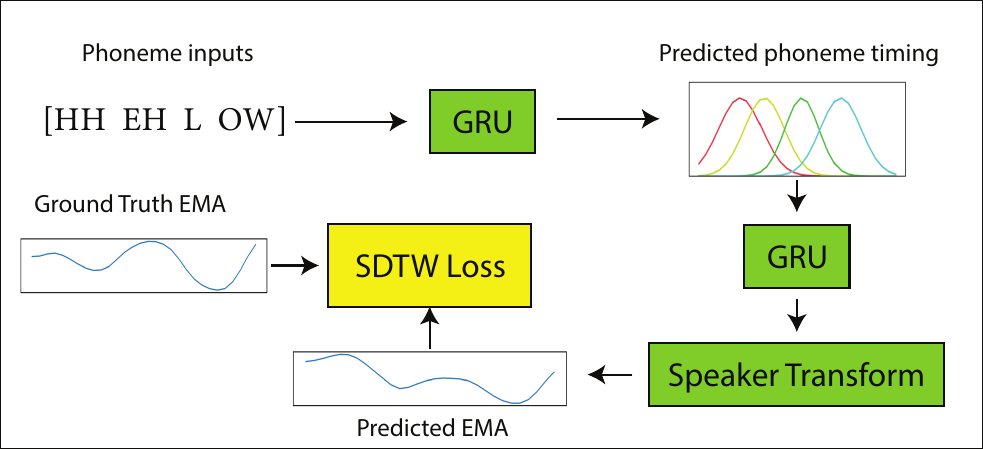}
  \caption{P2A model architecture}
  \label{fig:mode_arch}
\end{figure}

The P2A model generates an ``average'' expected EMA time series given any phonemic sequence input.
The model has three major components (Figure \ref{fig:mode_arch}): a bidirectional gated recurrent unit (GRU) \cite{chung2014empirical} neural network that predicts the timing of each phoneme, a second bidirectional GRU that generates speaker-independent EMA from the predicted phoneme timing, and a set of speaker-specific linear transforms that generate speaker dependent EMA. 
The entire model is trained end to end with a soft dynamic time warping (SDTW) \cite{cuturi2017soft} loss to account for utterance-specific timing variability that is not possible to predict from a phoneme sequence alone.
The code is available at~\url{https://github.com/cffan/ema_coarticulation}.

\subsection{Phoneme timing prediction}


Given an input phoneme sequence $\{\bm{x}_{1}, ..., \bm{x}_{n}  \}$, with each $\bm{x}_i \in \mathbb{R}^{c}$ representing a one-hot encoding of the $i$th phoneme, the phoneme timing bidirectional GRU model, denoted as $f_\theta$, predicts the influence of each phoneme at every timestep.
We use a Gaussian distribution to model the influence.
A similar idea was previously suggested by \cite{lofqvist1990speech}.
Phoneme $\bm{x}_i$'s Gaussian distribution is parameterized by two values, $\mu_i$ and $\sigma_i$, which are outputs of the $f_\theta$:
\begin{equation}
    \bm{M} = f_\theta (\{ \bm{x}_{1}, ..., \bm{x}_{n} \}) \quad \bm{M} \in \mathbb{R}^{n \times 2}
\end{equation}
where $\mu_i = M_{i,0}$ represents the time point of maximum influence for phoneme $i$, while $\sigma_i = M_{i,1}$ represents the width of its influence. 
The Gaussian model permits each time point to be dominated by one phoneme, with nearby Gaussians overlapping to produce coarticulation.
$\bm{M}$ is transformed into $\bm{M}' \in \mathbb{R}^{t \times n}$ representing the influence of $n$ input phonemes over $t$ timesteps  with a Gaussian kernel, i.e. $M_{k, i}' = \exp(-\frac{(k - \mu_{i})^2}{2 \sigma_i^2})$ 

\subsection{EMA generation}

Given the predicted timing $\bm{M}'$ of the input phonemes, our model proceeds to generate the final EMA time series $\bm{\hat{Y}}$ as follows:
\begin{equation}
    \bm{\hat{Y}} = h_\lambda (g_\phi (\bm{M}'))\quad \bm{\hat{Y}} \in \mathbb{R}^{t \times s}
\end{equation}
where $s$ is the number of EMA sensor measurements, $h_\lambda$ denotes a speaker-specific linear transform function, and $g_\phi$ is a bidirectional GRU that generates speaker-independent EMA representations.
Each row $\hat{Y}_{k,:}$ is the generated EMA measurements at time $k$.

\subsection{Model hyperparameters and training}

$f_\theta$ and $g_\phi$ are each defined as a 2-layer bidirectional GRU with 128 hidden units.
Each speaker has a unique speaker-specific linear transform function $h_\lambda$.
The smoothness parameter $\gamma$ in SDTW is set to 1. 
Hyperparameters are tuned on the validation set.
The P2A model is trained end-to-end using Adam optimizer with a learning rate of 4e-4 and batch size of 64 on one NVIDIA A100 GPU.
The gradient vector's L2 norm is scaled to 10 to stabilize training.


\section{Measuring Coarticulation}

To analyze coarticulation, we use minimal word pairs. These are pairs of words that differ in only one phoneme. For instance, ``pat \textipa{[p{\ae}t]}'' and ``bat \textipa{[b{\ae}t]}'' is a minimal pair that differs at the first phoneme. Coarticulation analysis has previously utilized minimal pairs of phonemic sequences \cite{gelfer1989determining,liu2022coarticulation}. We extend this method to larger sets of words.
More generally, given a minimal word pair that differs at position $i$, we have the following phonemic sequence:
\begin{equation*}
\label{eq:minimal_pair}
   [ ... P_{i-2} \, P_{i-1} \, \bm{P_{i}} \, P_{i+1} \, P_{i+2} ... ]
\end{equation*}
where $P_{i}$ is the coarticulation trigger, and other phonemes $P_{k|k \neq i}$ are the targets of coarticulation.
When comparing the difference between the EMA representations of phonemes at the same position in the pair, we expect the difference to peak at the $i$th position and gradually drop for further away targets.
We use the P2A model to generate EMA trajectories for the pair.
Then for each phoneme $P_{k}$ we time-average the EMA around its maximum influence timestep $\mu_k$ to obtain a vector representation $\bm{\varphi}_{k}$:
\begin{equation}
\label{eq:EMA_repr}
    \bm{\varphi}_{k} = \frac{1}{2\tau + 1} \sum_{j = \mu_k - \tau}^{\mu_k + \tau} \hat{Y}_{j,:} \quad \bm{\varphi}_{k} \in \mathbb{R}^{s}
\end{equation}
where $\tau$ is a hyperparameter controlling the width of the phoneme, and $s$ is the number of EMA sensor measurements.
We use $\tau=1$ for all experiments to minimize the influence of nearby phonemes.

We use Euclidean distance to measure the difference between each phoneme's EMA representation in two words \textit{A} and \textit{B} from a minimal pair:
\begin{equation}
\label{eq:ema_dist}
   d_k = || \bm{\varphi}^A_{k} - \bm{\varphi}^B_{k} ||
\end{equation}
By computing $d_k$ across all phonemic positions for many minimal word pairs, we can obtain a comprehensive and statistically significant view of modeled coarticulation magnitude and temporal extent.

\section{Results}



\subsection{Datasets}
We train the model on USC-TIMIT and Haskins Production Rate Comparison (HPRC) datasets.
USC-TIMIT is a 4-speaker dataset containing 1 hour of 100 Hz EMA. 
HPRC is an 8-speaker dataset containing 6.3 hours of 100 Hz EMA. 
Training on multiple speakers enables the model to learn average speaking patterns.
We use a $90\%$-$10\%$ train-validation split for USC-TIMIT, and $80\%$-$20\%$ split for HPRC.
Training features include the 2D positions (along the posterior-anterior and inferior-superior axes) and the 2D velocities of each EMA sensor.
Adding 2D velocities improves the quality of generated EMA.
Features are normalized by speaker-specific mean and standard deviation.
Only generated 2D positions are used to evaluate performance and measure coarticulation.
Given limited EMA training data, the phoneme input set is simplified by omitting stress markers from vowels, resulting in 39 phonemes and an additional silence phone. 

\subsection{Quality of generated EMA}

\begin{table}[h]
\caption{Comparing PCC and RMSE on model-predicted EMA (after dynamic time warping) and ground truth EMA on the USC-TIMIT and HPRC datasets. The confidence interval is computed via bootstrap resampling.}
\label{tab:ema_pred_acc}
    \centering
    \small
    \begin{tabular}{ccc}
    \toprule
    \textbf{Dataset} &    \textbf{PCC \tiny{[95\% CI]}}    & \textbf{RMSE \tiny{[95\% CI]}}        \\ \midrule
    USC-TIMIT       & 0.83 \,\tiny{[{0.82}, {0.84}]}                & 0.59  \,\tiny{[{0.57}, {0.62}]}     \\
    HPRC          & 0.85 \,\tiny{[{0.85}, {0.85}]}                &  0.53 \,\tiny{[{0.53}, {0.53}]}        \\ \bottomrule
    \end{tabular}
\end{table}

We compare the model-generated EMA with normalized ground truth EMA data using Pearson correlation coefficients (PCC) and root mean squared error (RMSE). 
The generated EMA data is aligned with the ground truth using dynamic time warping (DTW) before comparison.
Table \ref{tab:ema_pred_acc} shows the PCC and RMSE on two datasets.
To place these numbers in context, one SOTA A2A model achieves 0.78 PCC on the HPRC dataset \cite{wu23_icassp}, and an earlier model achieves 0.61 and 0.50 RMSE on USC-TIMIT and HPRC datasets \cite{parrot20_interspeech}, comparable to ours.
This shows that our P2A model can generate high-quality time-warped EMA.
We use data from speaker F01 in HPRC dataset for all the results below.

\setcounter{figure}{2}
\begin{figure*}[b]
  \centering
  \includegraphics[width=0.65\textwidth]{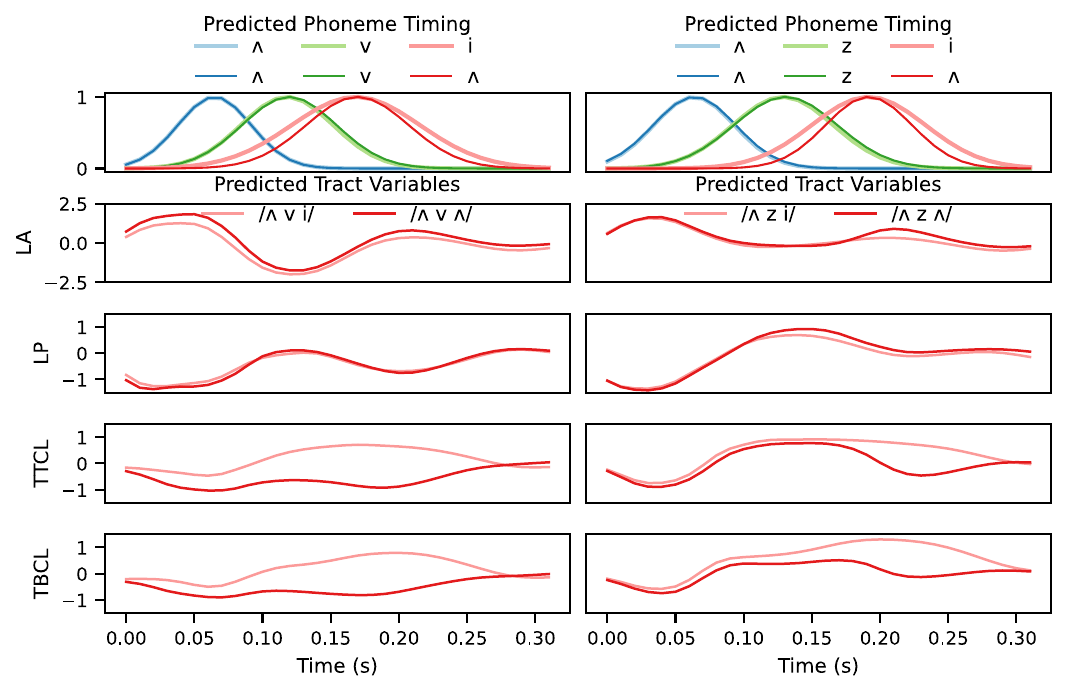}
  \caption{Model-generated EMA can reproduce known coarticulation on VCV pairs. \textbf{Left}: predicted phoneme influence and tract variables for \textipa{[2vi]} and  \textipa{[2v2]}, \textbf{Right}:\textipa{[2zi]} and \textipa{[2z2]}. TTCL and TBCL diverge before the onset of the last vowel, indicating anticipatory coarticulation. The divergence starts earlier in the left panel as compared to the right, showing the difference in coarticulation resistance between \textipa{[v]} and \textipa{[z]}.}
  \label{fig:coarticulation_eg}
\end{figure*}

\setcounter{figure}{1}
\begin{figure}[h]
  \centering
  \includegraphics[width=\linewidth]{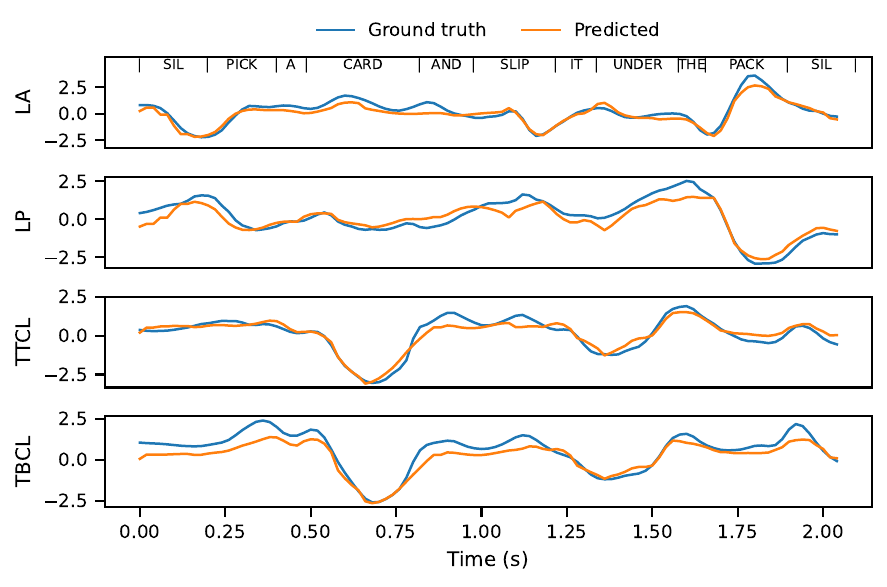}
  \caption{Comparing tract variables for model-generated vs. ground truth EMA for the sentence ``Pick a card and slip it under the pack''. Ground truth word alignments are shown in the top panel. The generated EMA is close to the ground truth EMA.}
  \label{fig:ema_pred_gt}
\end{figure}

Figure \ref{fig:ema_pred_gt} shows the generated EMA is qualitatively similar to the ground truth on a validation sentence.
Four normalized tract variables (TVs) are visualized: lip aperture (LA), lip protrusion (LP), tongue tip constriction location (TCCL), and tongue body constriction location (TBCL).
TVs are computed using lip and tongue sensor positions as in \cite{seneviratne19_interspeech}.

\subsection{Coarticulation examples}

We use two VCV pairs from \cite{fowler2000coarticulation} to show that our P2A model can generate EMA with known coarticulation patterns.
Figure~\ref{fig:coarticulation_eg} shows the predicted timing (1st row) and four normalized TVs (2nd to 5th row) for each pair.
The left panel compares \textipa{[2vi]} and \textipa{[2v2]}, and the right compares \textipa{[2zi]} and \textipa{[2z2]}.
The tongue position for \textipa{[i]} is more forward than that for \textipa{[2]}. 
As a result, the TTCL and TBCL for \textipa{[i]} (light red) in both panels have higher values, indicating a more frontal position, than those for \textipa{[2]} (dark red).
In both panels, the tongue tip and body positions start to diverge before the onset of the last vowel, showing anticipatory coarticulation.
The divergence starts earlier in the left panel than the right because \textipa{[z]} has to be produced with the tongue close to the alveolar ridge, whereas the tongue has less constraint for producing \textipa{[v]}.
This result is similar to that in \cite{fowler2000coarticulation}, indicating that our P2A model can accurately produce coarticulatory dynamics.

\subsection{Temporal extent and magnitude of coarticulation}

To comprehensively measure coarticulation, we enumerate 9,291 minimal word pairs from 10,000 most common words (according to Google Books NGram) in the CMU Pronunciation dictionary.
For each minimal pair, we use the P2A model to generate their EMA trajectories, extract each phoneme's EMA representation using Eq \ref{eq:EMA_repr}, and measure the EMA representation difference using Eq \ref{eq:ema_dist}.

Figure \ref{fig:coarticulation_extent} shows the temporal extent and magnitude of coarticulation.
The y-axis is the magnitude of coarticulation normalized to the average distance between all phonemes (such that a 100\% distance means that a phoneme changed so much as to resemble a completely different phoneme). 
The x-axis is the phonemic distance to the coarticulation trigger.
The most significant coarticulation effect (31\%$\pm$16\%, mean$\pm$std) occurs on the trigger's immediate left and right neighbors ($\pm$1). 
The coarticulation effect reduces significantly to 12\%$\pm$7\% for targets at $\pm$2 distance. 
For further targets, the coarticulation effect gradually decreases (8\% at $\pm$3 distance and 6\% at $\pm$4 distance).
The wide range of magnitude at each position is due to the varying distances between the trigger phonemes. For instance, trigger pairs such as \textipa{[S]} and \textipa{[\textteshlig]} cause less coarticulation than \textipa{[S]}~and~\textipa{[p]}.

We used random sentence pairs with either the first or last phonemes changed to measure the baseline magnitude due to model inaccuracy/noise at a far distance, shown as the dashed line.
Compared to the baseline, there is still a significant (p=0.01) coarticulation effect at $\pm$7 distance, indicating a longer-range coarticulation effect than has typically been found previously \cite{ahmed2019long,recasens1989long,west2000long}.
This result also provides a new justification for the triphone model \cite{jurafsky}, i.e., the majority of coarticulation can be modeled by considering only the left and right neighbors.

\setcounter{figure}{3}
\begin{figure}[]
  \centering
  \includegraphics[width=1\linewidth]{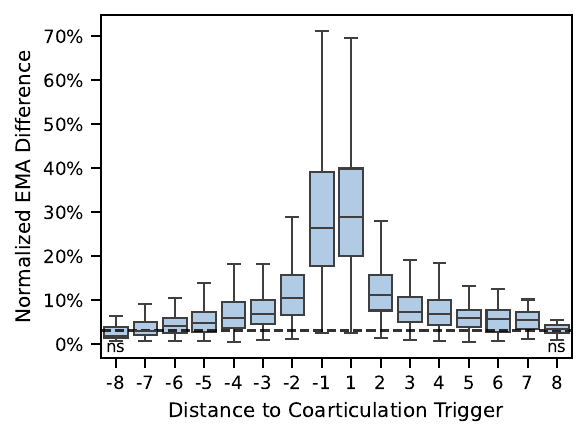}
  \caption{Quantifying the magnitude and extent of coarticulation. Each box shows the effects of coarticulation at a given phonemic distance to the trigger (whiskers mark the 1.5 interquartile range). The dashed line is the baseline EMA difference. Coarticulation is most pronounced at $\pm$1 distance, but is still significant compared to baseline at $\pm$7 distance.}
  \label{fig:coarticulation_extent}
\end{figure}

\subsection{Coarticulation resistance}
\begin{figure}[]
  \centering
  \includegraphics[width=0.8\linewidth]{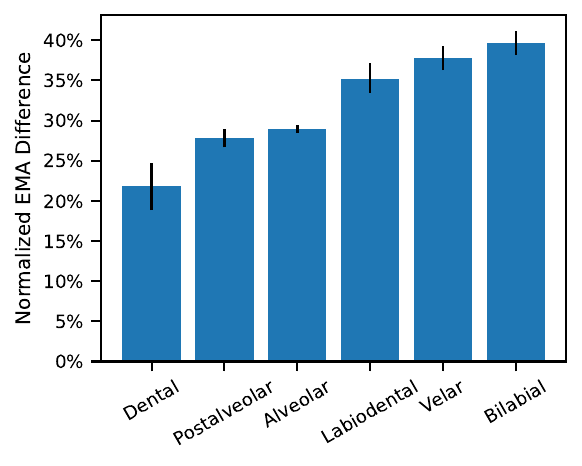}
  \caption{Comparing the coarticulation resistance of consonants (error bars represent bootstrap-resampled 95\% confidence intervals). Dentals, postalveolars, and alveolars are more resistant to coarticulation due to the involvement of the tongue.}
  \label{fig:coarticulation_resistance}
\end{figure}

\begin{minipage}{\columnwidth}
Coarticulation resistance refers to the degree to which phonemes resist being affected by adjacent phonemes. 
Our measure of coarticulation magnitude can be used as a measure of coarticulation resistance (smaller EMA differences indicate a greater resistance to coarticulation).
Figure \ref{fig:coarticulation_resistance} shows the effects of coarticulation on consonants, grouped by place of articulation, at $\pm$1 distance to the trigger.
Bilabials, labiodentals, and velars are more affected by coarticulation (lower coarticulation resistance) because producing bilabials and labiodentals does not require the tongue body, and velars can be produced with different tongue constriction locations \cite{dembowski1998articulator}.
In contrast, dentals, alveolars, and postalveolars produced with a raised and fronted tongue dorsum exhibit less coarticulation (higher coarticulation resistance). 
In one study \cite{fowler2000coarticulation} that analyzed coarticulation resistance of American English consonants, \textipa{[b]}, \textipa{[v]}, and \textipa{[g]} were found to be less resistant to coarticulation than \textipa{[D]}, \textipa{[d]}, \textipa{[Z]}, and \textipa{[z]}. 
These results are consistent with ours, suggesting that our framework can be flexibly used to study various aspects of \mbox{coarticulation}.
\end{minipage}

\section{Discussion}

Ideally, our findings would be validated using real EMA data. However, doing so is difficult, as it would require many repetitions of each word pair (to average over utterance-specific variability) and many word pairs. Searching over the HPRC dataset used in this work, we found only $\sim$60 word pairs per speaker that had at least 10 repetitions. Additionally, cross-word coarticulation and phoneme alignment errors in real EMA data could confound the results.
Alternatively, another future direction for validating this work could be to perform the same analysis on simulated acoustic data using advanced text-to-speech models, which aggregate thousands of hours of audio data \cite{wang2017tacotron,ren2020fastspeech}.

In future work, model-based analysis of coarticulation could provide new insight into coarticulation patterns in different languages, or into the development of coarticulation patterns in children. We intend to use the framework presented here to analyze neural data recorded during continuous speech \cite{willett2023high,metzger2023high}, which could provide insights into the neural mechanism of speech production.

This work is a first step towards a model-based, comprehensive analysis of coarticulation.
We leveraged existing EMA datasets and a novel P2A model to generate realistic EMA for many more minimal word pairs than would be realistically feasible for a typical coarticulation experiment.
Using the model-generated EMA, we measured the magnitude and temporal extent of coarticulation up to $\pm$8 phonemic distance from the trigger.
Our results are consistent with previous studies of coarticulation and suggest a more pronounced long-range coarticulation effect than previously appreciated.

\section{Acknowledgements}
We thank K. Livescu for her invaluable feedback and insightful comments, and B. Davis, K. Tsou, and S. Kosasih for administrative support. Support was provided by Wu Tsai Neurosciences Institute, Howard Hughes Medical Institute, Larry and Pamela Garlick, Simons Foundation Collaboration on the Global Brain, and NIH-NIDCD R01DC014034.

\bibliographystyle{IEEEtran}
\bibliography{mybib}

\end{document}